\documentclass[aps, prd, twocolumn, superscriptaddress, nofootinbib]{revtex4}
\usepackage{graphicx}
\usepackage{dcolumn}
\usepackage{amssymb}
\usepackage{subfigure}
\usepackage{array}
\usepackage[]{amsmath}
\usepackage{epstopdf}
\usepackage[usenames]{color}

\newcommand{\bv}[1]{\boldsymbol{#1}}

\newcommand{\be}{\begin{equation}}
\newcommand{\ee}{\end{equation}}
\newcommand{\bea}{\begin{eqnarray}}
\newcommand{\eea}{\end{eqnarray}}

\def\dprime{\mathaccent"707D}
 
\newcommand{\ben}{\begin{equation}}
\newcommand{\een}{\end{equation}}
\newcommand{\ba}{\begin{array}}
\newcommand{\ea}{\end{array}}
\newcommand{\bit}{\begin{itemize}}
\newcommand{\eit}{\end{itemize}}

\def\math{\mathsurround 0pt}
\def\oversim#1#2{\lower.5pt\vbox{\baselineskip0pt \lineskip-.5pt
    \ialign{$\math#1\hfil##\hfil$\crcr#2\crcr{\scriptstyle\sim}\crcr}}}

\definecolor{Blue}{rgb}{0.0,0,1.0} 
\definecolor{Green}{rgb}{0.0,0.8,0.0} 
\definecolor{MyDarkBlue}{rgb}{0.1,0,0.55}

\begin{document}
\title{Particle motion in weak relativistic gravitational fields}

\newcommand{\addressSussex}{Department of Physics \&
Astronomy, University of Sussex, Brighton, BN1 9QH, United Kingdom}

\newcommand{\addressCAP}
{
D\'epartement de Physique Th\'eorique and Center for Astroparticle Physics, Universit\'e de Gen\`eve, Quai E.\ Ansermet 24, CH-1211 Gen\`eve 4, Switzerland
}

\author{Miki Obradovic}
\email{miki.obradovic@unige.ch}
\affiliation{\addressCAP}
\affiliation{\addressSussex}

\author{Martin Kunz}
\email{martin.kunz@unige.ch}
\affiliation{\addressCAP}

\author{Mark Hindmarsh} 
\email{m.b.hindmarsh@sussex.ac.uk}
\affiliation{\addressSussex}

\author{Ilian T. Iliev}
\email{I.T.Iliev@sussex.ac.uk}
\affiliation{\addressSussex}

\date{November 20, 2012}

\begin{abstract}
We derive the geodesic equation of motion in the presence of weak gravitational fields produced
by relativistic sources such as cosmic strings, decomposed into scalar, vector and tensor parts.
To test the result, we perform the first N-body simulations with relativistic weak gravitational 
external fields.  Our test case is a moving straight string, for which we recover the well-known result for the 
impulse on non-relativistic particles. We find that the vector (gravito-magnetic) force is an essential contributor. 
Our results mean that it is now possible to incorporate straightforwardly into N-body simulations all weak relativistic sources, 
including networks of cosmic defects. 
\end{abstract}

\maketitle

\section{Introduction} \label{sec:introduction}

Topological defects such as cosmic strings \cite{VilenkinShellardBOOK,Hindmarsh:1994re,Sakellariadou:2006qs,Copeland:2009ga} are generic by-products of many inflationary models \cite{Yokoyama:1989pa,Linde:1993cn,Copeland:1994vg,Sakellariadou:2007bv} and of Grand Unification  \cite{Jeannerot:2003qv}, adding a characteristic signature to the gravitational and matter fluctuations predicted by inflation.
Precision Cosmic Microwave Background (CMB) measurements set limits on the 
allowed defect abundance \cite{Bevis:2004wk,Bevis:2007gh,Urrestilla:2007sf,Hindmarsh:2011qj,Battye:2006pk,Battye:2010hg,Battye:2010xz,Landriau:2002fx} 
and thus also on inflationary models that produce defects.

There is also accurate data on the galaxy power spectrum over a wide range
of scales \cite{Reid:2009xm}, which constrain the matter perturbations.
Inflation creates a nearly Gaussian spectrum of initial perturbations 
(see e.g. \cite{Lyth:2009zz}), whose subsequent evolution can be computed
in linear theory, and compared to the data (under assumptions about the
bias, i.e. the ratio between the galaxy and matter power spectra).
However, defects are localised and ``active" \cite{AlbrechtPhysRevLett.68.2121}
sources of gravitational fields, creating highly non-Gaussian perturbations
\cite{Hindmarsh:2009qk,Hindmarsh:2009es,Ringeval:2010ca}.  For example, cosmic
strings create a wake behind them as they move through matter  
\cite{1987ApJ...322....1S,PTP.77.1152}, in which there is a planar
relative overdensity of order 1 as soon as it is created. The evolution
of the wake is therefore immediately non-linear, and standard linear 
theory in Fourier space does not apply. 
The strong non-Gaussianity is very likely to impact
the growth of structure and (for example) could affect the bias of the galaxy power spectrum. 
For this reason, we need to find ways to go beyond linear perturbation theory to calculate
the matter power spectrum derived from the gravitational perturbations of defects. 

One way to do so involves N-body simulations. In some early numerical work 
by \cite{1997ApJ...482...22S,1997PhRvD..56.6139S} the structure of the wake 
induced by a single straight string was studied by setting up a velocity kick
towards the plane behind the string as the initial velocity perturbation, as
derived from semi-analytical predictions \cite{PhysRevLett.53.1700,VilenkinShellardBOOK}, 
verifying the predicted width of the wake, the inflow velocity of the dark matter and 
assumptions about the self-similarity of the solution. 

However, in general defects have a complex and evolving 3-dimensional structure
that extends up to the horizon scale. 
For example, strings are not straight but form a tangled, self-intersecting 
and fast-moving network of infinite and closed pieces, with a characteristic 
length scale of about one third of the horizon scale and a characteristic speed
of about a half that of light \cite{Hindmarsh:2008dw,Moore:2001px}.

In order to capture the full impact of the defect perturbations on the large-scale
structure, we will need to work on cosmological scales and with relativistic sources. 
For this we will need general relativity, both for the perturbations in the
gravitational field, and the deviations to the motion that the field produces.

In this paper we derive from first principles the equation of motion (EOM) of massive
particles in a perturbed Friedmann-Lema\^\i tre-Robertson-Walker (FLRW) cosmology,
keeping all terms linear in the gravitational fields, including vector and tensor.
This is required for relativistic sources such as topological defects, where all
parts of the energy-momentum tensor are comparable in magnitude.
We will then test our formalism with the help of a moving straight string, for
which the impulse on passing particles is known exactly. 

Finally, we perform an N-body simulation
to study the growth of the wake behind the moving
string, comparing with previous work using Newtonian
gravity, and the impulse as an 
initial condition \cite{1997ApJ...482...22S,1997PhRvD..56.6139S}.

This paper establishes the formalism by which the effects of topological defects 
on the matter in the universe can be taken into account, as a necessary preliminary
to non-perturbative (especially N-body) calculations of the growth of structure in
cosmological models with defects. The formalism is more general, however, and allows
to add any sources of weak relativistic gravitational fields to N-body simulations.

\section{Overview of linear perturbation theory}

Very generally, given a metric $g_{\mu\nu}$ we can compute the Christoffel symbols
\begin{equation} \label{eq:CrMetric}
\Gamma^\mu_{\alpha\beta} = 
\frac{1}{2}
g^{\mu\nu}
\left(
\partial_\alpha g_{\beta\nu} +
\partial_\beta g_{\alpha\nu} -
\partial_\nu g_{\alpha\beta}
\right) \, ,
\end{equation}
the curvature tensor $R_{\mu\nu}$ and thus the Einstein tensor $G_{\mu\nu}$ as well as the
geodesic equation of motion,
\begin{equation} \label{eq:geom}
\frac{d^2x^{\mu}}{d\tau^2} + \Gamma^{\mu}_{\alpha\beta} [g_{\rho \sigma}] \frac{dx^{\alpha}}{d\tau}\frac{dx^{\beta}}{d\tau} = 0 .
\end{equation}
The gravitational field equations or Einstein equations,
\begin{equation}
G_{\mu\nu} = 8 \pi G T _{\mu\nu}  \, , 
\end{equation}
describe the interaction between gravity and matter, with the latter given by its energy-momentum
tensor $T_{\mu\nu}$.

In general the Einstein equations are very difficult to solve, but in the limit of weak fields and small perturbations
we can use linear perturbation theory around a fixed background metric \cite{Kodama:1985bj,Mukhanov:1990me,Ma:1995ey}.
These linearised equations then naturally decompose into irreducible components under rotations, scalars (S), vectors
(V) and tensors (T). However, at higher order in perturbation theory this decomposition is no longer maintained
and the different types of perturbations mix. 

We will assume throughout this paper that the sources of the gravitational perturbations evolve on the background
universe and are not affected by the perturbations that they generate, i.e. that their perturbations 
only affect the remaining constituents. This is the case for topological defects, for which we can
perform the numerical simulations separately, recording the gravitational perturbations which are then self-consistently
inserted into the linearised Einstein equations of the full system \cite{Durrer1994FCPh...15..209D}. By running the defect
and the N-body simulation in parallel and exchanging information between the two we could in principle
include the so-called gravitational backreaction on the defects, but this is left for later work.

\subsection{Metric and Christoffel symbols}\label{ap:metric}

We choose a background metric $a^2 \eta_{\mu\nu}$, where $\eta_{\mu\nu} =  {\rm diag}(-1,1,1,1)$ is the
Minkowski metric (i.e. we only consider flat space) and a perturbation $a^2 h_{\mu\nu}$
so that the full metric with the S,V,T decomposition becomes
\begin{equation}
\begin{array}{cc}
g_{\mu\nu} = a^2(\eta_{\mu\nu} + h^{(S)}_{\mu\nu} + h^{(V)}_{\mu\nu} + h^{(T)}_{\mu\nu})
\end{array}
\end{equation}

The scalar perturbations in the conformal Newtonian gauge, the vector perturbations in the \textit{vector} gauge and the gauge invariant tensor perturbations are defined by
\begin{equation}
h^{(S)}_{\mu\nu}dx^\mu dx^\nu = -2 \psi d\tau^2 + 2 \phi \delta_{ij} dx^i dx^j
\end{equation}

\begin{equation}
h^{(V)}_{\mu\nu}dx^\mu dx^\nu = -2\Sigma_i d\tau dx^i \qquad
k^i\Sigma_i = 0 
\end{equation}
\begin{equation}
h^{(T)}_{\mu\nu}dx^\mu dx^\nu = h^{(T)}_{ij} dx^i dx^j \qquad
h^{(T)}_{ij}k^j = h{^{(T)i}_i} = 0 
\end{equation}

where $\tau$ is conformal time.

From the metric we can immediately derive the Christoffel symbols with the help of Equation (\ref{eq:CrMetric}).
Using primes ($\acute \; \equiv \partial_{\tau}$) to denote derivatives with respect to conformal time, we find
\begin{eqnarray}
\Gamma^\mu_{\alpha\beta} &=&
\frac{1}{2a^2}
\Bigg[
-\left(h^{(S)}_{\mu\nu} - h^{(V)}_{\mu\nu} + h^{(T)}_{\mu\nu}\right)
\nonumber\\&& \times
\left(\partial_\beta a^2 \eta_{\alpha\nu} 
\partial_\alpha a^2 \eta_{\beta\nu} - 
\partial_\nu a^2 \eta_{\alpha\beta}
\right)
\nonumber\\&& 
+\, \eta^{\mu\nu}
\Big(
\partial_\beta \left[a^2(\eta_{\alpha\nu} + h^{(S)}_{\alpha\nu} + h^{(V)}_{\alpha\nu} +h^{(T)}_{\alpha\nu})\right]
\nonumber\\&&
+\,\partial_\alpha\left[a^2(\eta_{\beta\nu} + h^{(S)}_{\beta\nu} + h^{(V)}_{\beta\nu} +h^{(T)}_{\beta\nu})\right]
\nonumber\\&&
-\, \partial_\nu \left[a^2(\eta_{\alpha\beta} + h^{(S)}_{\alpha\beta} + h^{(V)}_{\alpha\beta} +h^{(T)}_{\alpha\beta})\right]
\Big) 
\Bigg] .
\label{}
\end{eqnarray}

In particular, the zeroth component is found to be
\begin{eqnarray}
\Gamma^0_{\alpha\beta} &=&
\frac{\acute a}{a}
\left[
2 \delta^0_\alpha\delta^0_\beta
+ \eta_{\alpha\beta} + h^{(S)}_{\alpha\beta} + h^{(V)}_{\alpha\beta} + h^{(T)}_{\alpha\beta} - 2\psi\eta_{\alpha\beta} 
\right]
\nonumber \\&&
+ 
\frac{1}{2}
\big[
2\partial_\beta\psi\delta^0_\alpha 
+\partial_\beta\Sigma_i\delta^i_\alpha
+ 2 \partial_\alpha \psi \delta^0_\beta
+ \partial_\alpha\Sigma_i\delta^i_\beta 
\nonumber \\&&
 +\partial_0(h^{(S)}_{\alpha\beta} + h^{(V)}_{\alpha\beta} + h^{(T)}_{\alpha\beta})
\big]
\end{eqnarray}
and the i'th component is found to be
\begin{eqnarray}
\Gamma^i_{\alpha\beta} &=& \nonumber
\frac{\acute a}{a} 
\left[
\delta^i_\alpha\delta^0_\beta + \delta^i_\beta\delta^0_\alpha
+ \Sigma_i(\delta_{ij}\delta^i_\alpha\delta^j_\beta - \delta^0_\alpha\delta^0_\beta )
\right]
 \\ \nonumber &&
- \frac{1}{2} \partial_i(h^{(S)}_{\alpha\beta} + h^{(V)}_{\alpha\beta} + h^{(T)}_{\alpha\beta})
+ \acute \phi (\delta^i_\alpha\delta^0_\beta + \delta^i_\beta\delta^0_\alpha)
\\ \nonumber & &
+ \partial_j\phi(\delta^i_\alpha\delta^j_\beta + \delta^i_\beta\delta^j_\alpha) 
- \acute\Sigma_i \delta^0_\alpha\delta^0_\beta
\\ \nonumber &&
- \frac{1}{2} \partial_j\Sigma_i(\delta^0_\alpha\delta^j_\beta + \delta^0_\beta\delta^j_\alpha) 
+ \frac{1}{2} \acute h^{(T)}_{ji}(\delta^j_\alpha\delta^0_\beta + \delta^j_\beta\delta^0_\alpha)
\\&&
+ \frac{1}{2}  \partial_k h^{(T)}_{ji}(\delta^j_\alpha\delta^k_\beta + \delta^j_\beta\delta^k_\alpha) .
\end{eqnarray}

\subsection{Einstein Equations}

We solve the linearized Einstein equations in Fourier space where 
they have a simpler form. Our Fourier transform conventions are

\begin{equation}
f(k) = \int_{-\infty}^{\infty} f(x) e^{i k x} dx \qquad
f(x)  =  \frac{1}{2\pi}\int_{-\infty}^{\infty} f(k) e^{-i k x} dk 
~.
\end{equation}

Following the formalism of \cite{Durrer:2001cg}, the Einstein
equations in Fourier space can be written as

\begin{eqnarray}\label{eq:EINSTEIN}
\phi = \frac{4\pi G}{k^2} (f_{\rho} + 3\frac{\acute a}{a}f_v) \qquad
\psi = -8\pi G f_{\pi} - \phi  \\
\Sigma_i =  - \frac{16 \pi G}{k^2}w_i^{(V)} \qquad \nonumber
\dprime h^{(T)}_{ij} + 2\frac{\acute a}{a} \acute h^{(T)}_{ij} +k^2 h^{(T)}_{ij} &=& 8\pi G\tau_{ij}^{(\pi)}
\end{eqnarray}

In these expressions we used the following elements of the energy momentum tensor $T_{\mu\nu}$:

\begin{eqnarray}\label{eq:Tprojectionoperators}
f_{\rho} = T_{00}  \qquad
f_{v} = \frac{i \hat{k}_j}{k} T_{0}^j \qquad
f_{p} = \frac{1}{3}\delta_{ij} T^{ij} \qquad \nonumber \\
f_{\pi} = -\frac{3}{2k^2}(\hat{k}^i\hat{k}^j-\frac{1}{3}\delta_{ij} )T_{ij} \nonumber \qquad
w_i^{(V)} = (T_{0i} - \hat k_i \hat k^j T_{0j}) \nonumber \\ 
\tau_{ij}^{(\pi)} = \left(P_{il}P_{jm}-(1/2)P_{ij}P_{lm}\right)P^{ma}P^{lb}T_{ab} \qquad
\end{eqnarray}
where hats denote unit vectors, and where we used the projection operator
\begin{equation}
P_{ij} = \delta_{ij} - \hat k_i \hat k_j ~.
\end{equation}

\section{Particle motion in weak gravitational fields}

Evaluating the zeroth component of the geodesic equation (\ref{eq:geom}) gives us an expression for the evolution of the energy of massive particles

\begin{eqnarray}
\frac{\acute E}{E} &=&
- \acute\psi - 2\partial_j\psi\acute x^j -\acute\phi\acute x_j\acute x^j - (\partial_j\Sigma_i + \frac{1}{2}\acute h^{(T)}_{ij})\acute x^i \acute x^j
\nonumber \\ &&
- \frac{\acute a}{a}
\left[
(1 - 2 \psi + 2 \phi)\acute x_j \acute x^j 
-2 \Sigma_i \acute x^i
+ h^{(T)}_{ij}\acute x^i \acute x ^j
\right] \nonumber ~. \\
\end{eqnarray}

Using this in the $i$ equation we find the equation of motion of massive particles in
a weak gravitational field:

\begin{widetext}
\begin{eqnarray}\label{eq:EOMfullyperturbed}
\dprime x^i &=&  
\left[
\acute \psi - 2\acute\phi + 2 \partial_j(\psi-\phi)\acute x^j 
+ \acute \phi \acute x_j \acute x^j + (\partial_j\Sigma_k + \frac{1}{2}\acute h^{(T)}_{ij})\acute x^k\acute x^j
\right]\acute x^i 
\nonumber
+ \frac{\acute a}{a} 
\left[
- 1 + (1 - 2\psi + 2\phi) \acute x_j \acute x^j - 2\Sigma_j\acute x^j + h^{(T)}_{kj}\acute x_k \acute x^j
\right]\acute x^i \\ &&  
- \partial_i(\psi -\phi\acute x_j \acute x^j)
+ \acute \Sigma_i + \partial_j\Sigma_i\acute x^j 
- \partial_i\Sigma_j \acute x^j
- \frac{\acute a}{a}\Sigma_i(\acute x_j \acute x^j -1)
- \acute h^{(T)}_{ji}\acute x^j - \partial_k h^{(T)}_{ji}\acute x^j \acute x^k 
 + \frac{1}{2}\partial_i h^{(T)}_{kj}\acute x^k \acute x^j ~.
\end{eqnarray}
\end{widetext}

We notice that all types of perturbations, scalar, vector and tensor, affect the particle motion. However,
to leading order for non-relativistic particles with $\acute{x}\ll 1$ 
we find that tensor perturbations do not contribute. 

Converting (\ref{eq:EOMfullyperturbed}) to physical time and retaining only the leading order terms in the particle velocity we find 
\begin{equation}\label{eq:EOMfullyperturbedzerothorder}
\ddot x^i 
+ 2\frac{\dot a}{a} \dot x^i
= 
-\frac{1}{a^2}\partial_i\psi 
+ \frac{1}{a}\dot \Sigma_i 
+ \frac{\dot a}{a^2} \Sigma_i ~ .
\end{equation}
The second term on the left is due to the expansion of space. Although it contains $\dot{x}$ it needs to be taken into account as it is not suppressed by any of the metric perturbations.
The first term on the right is the usual gradient
of the gravitational potential which is the sum of the potential due to the particles and the scalar perturbation
potential of the topological defects.\footnote{We allow
for a non-zero anisotropic stress since for topological defects in general $\phi \neq \psi$. We note that only
the $\psi$ potential accelerates massive particles to lowest order in the particle velocity.} However,
we find that vector-type metric perturbations, if they are present, affect particle motion at the same level. Thus gravito-magnetic forces can be as important as the standard scalar force for a relativistic source.
This expression should be used when extending N-body codes to account for general sources of weak gravitational fields.
The full result needs to be evolved if particles can reach relativistic speeds, and it may be worth to occasionally monitor
the size of the next order terms.

\section{Particle motion induced by a straight string in 
Minkowski space}

\begin{figure}[!htb]
\subfigure{\includegraphics[scale=0.7]{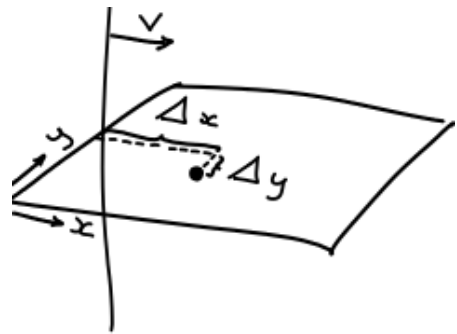}}
\caption{A Nambu Goto string is aligned with 
the $z$ axis and is travelling in the $x$ direction at constant 
speed $v$ through the middle of the $xy$ plane. A test 
particle is at a position $(\Delta x, \Delta y)$ w.r.t. 
the initial position of the string such
that $\Delta y / \Delta x \ll 1$. The string travels 
a distance $2 \Delta x$.}
\label{fig:sketch}
\end{figure}

In this section we will reproduce the well known effect 
of a velocity kick induced by a moving straight Nambu-Goto 
string on a test particle, in a non-expanding Minkowski 
background with $a\equiv 1$ (see Fig. \ref{fig:sketch}). We compute the result by 
evolving equation (\ref{eq:EOMfullyperturbedzerothorder}), 
which illustrates the approach that we will be using in 
the future for the full network and verifies that we arrive 
at the correct answer.
We show that both scalar and vector perturbations contribute 
significantly to the particle motion, even though in the 
approximation where the particle is scattered by a string 
coming from infinity, in the infinite time limit only the 
scalar contribution remains relevant.

The string Nambu Goto action is given by
\begin{equation}
S = -\mu \int \sqrt{-g^{(2)}}d^2\zeta
\end{equation}
where $\mu$ is the string mass density, $\zeta^\alpha$ ($\alpha = 0,1$) are the 
coordinates on the worldsheet traced out by the string with spacetime coordinates $X^\mu(\zeta)$,
and $g^{(2)}$ is the determinant of the induced metric on the world-sheet,  $g^{(2)}_{\alpha\beta} = \partial_\alpha X\cdot\partial_\beta X$.

The particular gauge chosen for defining the worldsheet
is \cite{VilenkinShellardBOOK}
\begin{equation}\label{eq:string_fp_0}
g^{(2)}_{01} = 0, \qquad
g^{(2)}_{00} + g^{(2)}_{11} = 0,
\end{equation}
and we may also identify worldsheet and coordinate time with the choice
$\zeta^0 = t$.

We are making the simplification of the string living in an 
flat, non-expanding space. Its energy momentum tensor is 
\cite{VilenkinShellardBOOK}
\begin{equation}\label{eq:stringfpEMT}
T^{\mu\nu}(\bv{x},t) = 
\mu \int d\lambda 
\left(
\dot{X}^\mu\dot{X}^\nu 
- \acute{X}^\mu\acute{X}^\nu
\right)
\delta^{(3)}\left(\bv{x} -\bv{X}(\lambda,t)\right)
\end{equation}
where $\dot{X}^\mu \equiv \partial_0X^\mu$ and $\acute{X}^\mu \equiv \partial_1 X^\mu$.

We are considering specifically a straight string parallel to the 
$z$ axis, traveling in the $x$ direction at a constant 
velocity $v$ \footnote{$v$ in units of $c=1$, $\gamma$ 
is the Lorentz factor.} so that
\begin{equation}\label{eq:stringfp3}
\begin{array}{cc}
\dot{X}^\mu = (1,v,0,0)
, &
\acute{X}^\mu = (0,0,0,\frac{1}{\gamma}),
\end{array}
\end{equation}
In this case we find that the energy momentum tensor is given by
\begin{equation}
T^{\mu\nu}(\bv{x},t) =
\gamma \mu
M^{\mu\nu}
\delta(x - x_0 - vt)
\delta(y - y_0)
\end{equation}
with 
\begin{equation}
M^{\mu\nu} =
\left(
\begin{array}{cccc}
1 & v & 0 & 0 \\
v & v^2 & 0 & 0 \\
0 & 0 & 0 & 0 \\
0 & 0 & 0 & -1/\gamma^2
\end{array}
\right) ~.
\end{equation}

We now change to Fourier space in order to solve for 
the perturbations in the metric with the help of the 
Einstein equations:
\begin{equation}
T^{\mu\nu}(\bv{k},t) =
2 \pi
\gamma \mu
M^{\mu\nu}
e^{i k_x (x_0 + vt)}
e^{i k_y y_0}
\delta(k_z)
\end{equation}

Using equations (\ref{eq:EINSTEIN}) with $a=1$ we find the perturbations to be
(for a detailed calculation of Equation (\ref{eq:TensSolFourierMink}) 
see Appendix \ref{ap:tensorperturbations})

\begin{equation}
\phi(\bv{k},t) = \frac{8\pi^2 G \mu \gamma}{k^2}  e^{i k_x (x_0 + vt)} e^{i k_y y_0} \delta(k_z) 
\label{eq:APsikt}
\end{equation}
\begin{equation}
\psi(\bv{k},t) = \frac{8\pi^2 G \mu \gamma}{k^2}
v^2 \left(3\hat{k}_x^2 - 2\right) e^{i k_x (x_0 + vt)} e^{i k_y y_0} \delta(k_z) 
\label{eq:gaugeinvariantvectorfield}
\end{equation}
\begin{equation}
\Sigma_i(\bv{k},t) = \frac{32 \pi^2 G \mu \gamma }{k^2} \left( v_i - \hat k_i \hat k_x v \right)  e^{i k_x(x_0 + v t)}e^{i k_y y_0}\delta(k_z) 
\end{equation}
\begin{equation}
h^{(T)}_{ij} = \frac{8\pi G\tau_{ij}^{(\pi)}}{k^2 - k_x^2v^2}
\label{eq:TensSolFourierMink}
\end{equation}

In our setup, we have that $v_i = (v,0,0)$.
To calculate the effect of the perturbations on a particle 
to first order (Equation \ref{eq:EOMfullyperturbedzerothorder}) 
we only need $\psi$ and $\Sigma_i$. Their inverse Fourier 
transform is given by
\begin{eqnarray}
\psi({\bv{x},t}) &=& \frac{G\mu \gamma v^2}{2r^2}
\left[  -6x^2 + r^2
\log \left( \frac{r^2}
{r_0^2} \right)   \right] \label{eq:APsiXt} \\
\Sigma_1({\bv{x},t}) &=&  \frac{2G\mu\gamma v}{r^2} \label{eq:sigma_1_x}
\left[  2x^2 - r^2
\log \left( \frac{r^2}
{r_0^2} \right)   \right] \label{eq:gaugeinvariantvectorfieldXt} \\
\Sigma_2({\bv{x},t}) &=&  \frac{4G\mu\gamma v x y }{r^2} \label{eq:sigma_2_x} 
\\
\Sigma_3({\bv{x},t}) &=& 0 \label{eq:givfXt3} \label{eq:sigma_3_x}
\end{eqnarray}
where $r_0$ is an integration constant that sets the distance 
at which the logarithmic contribution of the infinite straight 
string to $\psi$ and $\Sigma_1$ vanishes, and 
$(x,y) =(x_0 + v t,y_0)$. Redefining the variables
\bea
x &\rightarrow& r_x =  x - x_0 - v t \\
y &\rightarrow& r_y = y - y_0
\eea
equations (\ref{eq:APsiXt}) -- (\ref{eq:givfXt3}) give the 
fields at position $(x,y)$ due to a string at $(x_0 + v t,y_0)$. 
Inserting these expressions into equation~(\ref{eq:EOMfullyperturbedzerothorder}) we arrive, after ignoring collisions where the 
solution is divergent, at the following equation of motion 
for particles:

\begin{equation}
\ddot x = \frac{G\mu\gamma v^2}{r^4}  (3r_x^2 + r_y^2) r_x 
\label{eq:EOMexplicitFullyperturbed_ax} 
\end{equation}
\begin{equation}
\ddot y = - \frac{G\mu\gamma v^2}{r^4} 
(3 r_x^2 + 5r_y^2) r_y 
\label{eq:EOMexplicitFullyperturbed_ay}
\end{equation}

\begin{figure}
\begin{center}    
\subfigure{\includegraphics[scale=1]{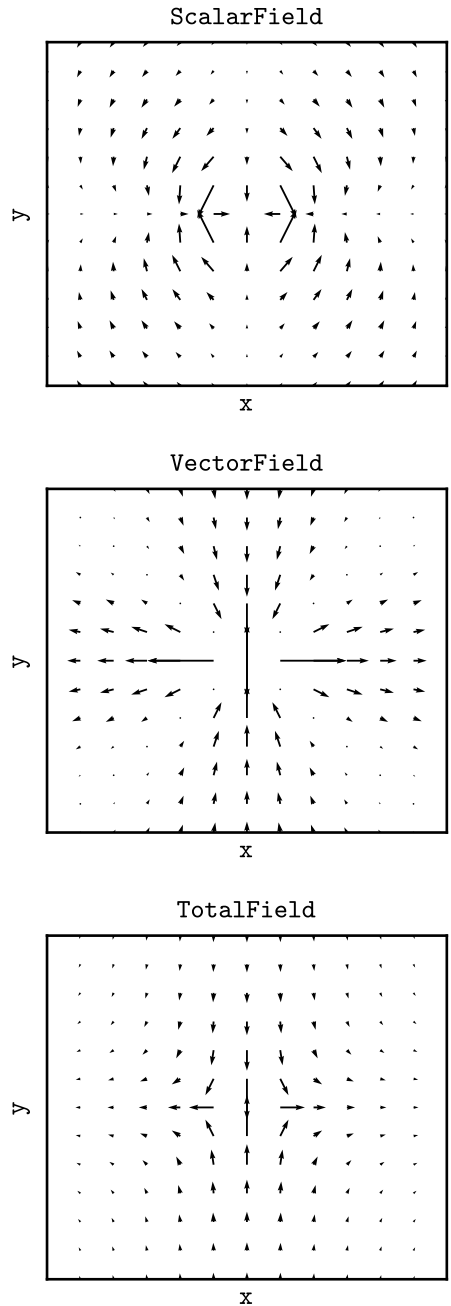}}
\end{center}    
\caption{The acceleration field around a straight string in Minkowski space-time moving at constant speed in the x direction (i.e. horizontally through the center of the box). The top panel shows only the contribution from the scalar components, and the middle panel from the vector components of the metric perturbations induced by the string. The bottom panel shows the total acceleration field which is the one seen by a particle.}
\label{fig:PACCELERATIOLNFIELD}
\end{figure}

We notice that, as expected, the logarithmically divergent 
part of the scalar and vector potentials $\psi$ and $\Sigma_i$ 
have been removed by the derivatives and the acceleration is 
decaying like $1/r$ at large distances. The acceleration field 
around a straight string is shown in 
Figure~\ref{fig:PACCELERATIOLNFIELD}. 

We solve the equations for $x(t)$ and $y(t)$ numerically with 
a standard numerical solver\footnote{NDSolve of the Mathematica software package.} for ordinary differential equations 
and for initial conditions chosen so that the string passes close to 
a test particle in the $y$ direction (at a distance $\Delta y$), 
starting at a distance $\Delta x$ far away and moving past the 
particle in the $x$ direction again to a distance $\Delta x$ 
such that $\Delta y / \Delta x \ll 1$ (the near-by limit). We 
find that the net acceleration in the $x$ direction cancels 
roughly out, while the net effect in the $y$ direction, in 
particular the \textit{velocity kick}, agrees with the predicted 
value \cite{VilenkinShellardBOOK} 
\begin{equation}\label{eq:predictedvelocitykick}
    u_{\mathrm{p}} = 4 \pi G \mu \gamma v .
\end{equation}
Our numerical result for the effect of a GUT string with 
$G \mu \approx 10^{-6}$ travelling at $v=0.333$ past a 
particle with ${\Delta y}/{\Delta x} \approx 0.002$ is
\begin{equation}
    u/u_{\mathrm{p}} \approx 0.9993 \, ,
\end{equation}
and the agreement can be improved nearly arbitrarily as 
${\Delta y}/{\Delta x} \rightarrow 0$. We also inserted the
modified acceleration 
equations~(\ref{eq:EOMexplicitFullyperturbed_ax}) and 
(\ref{eq:EOMexplicitFullyperturbed_ay}) into the 
public N-body code Gadget-2 \cite{Springel:2005mi}
and found that the N-body code result agrees with the result from the numerical 
solver to machine accuracy as long as as we resolve the dynamical 
time of the particle string interaction in the N-body code. This 
can be tuned to any precision when looking at the effect on single 
particles (however, this may be a problem to be solved when doing 
large scale N-body simulations where a general small limit on the 
maximum timestep is too expensive). We conclude that our 
formalism and simulation set-up reproduces the standard 
results in non-expanding space time accurately.

\begin{figure*}
\begin{center}    
\subfigure{\includegraphics[scale=1]{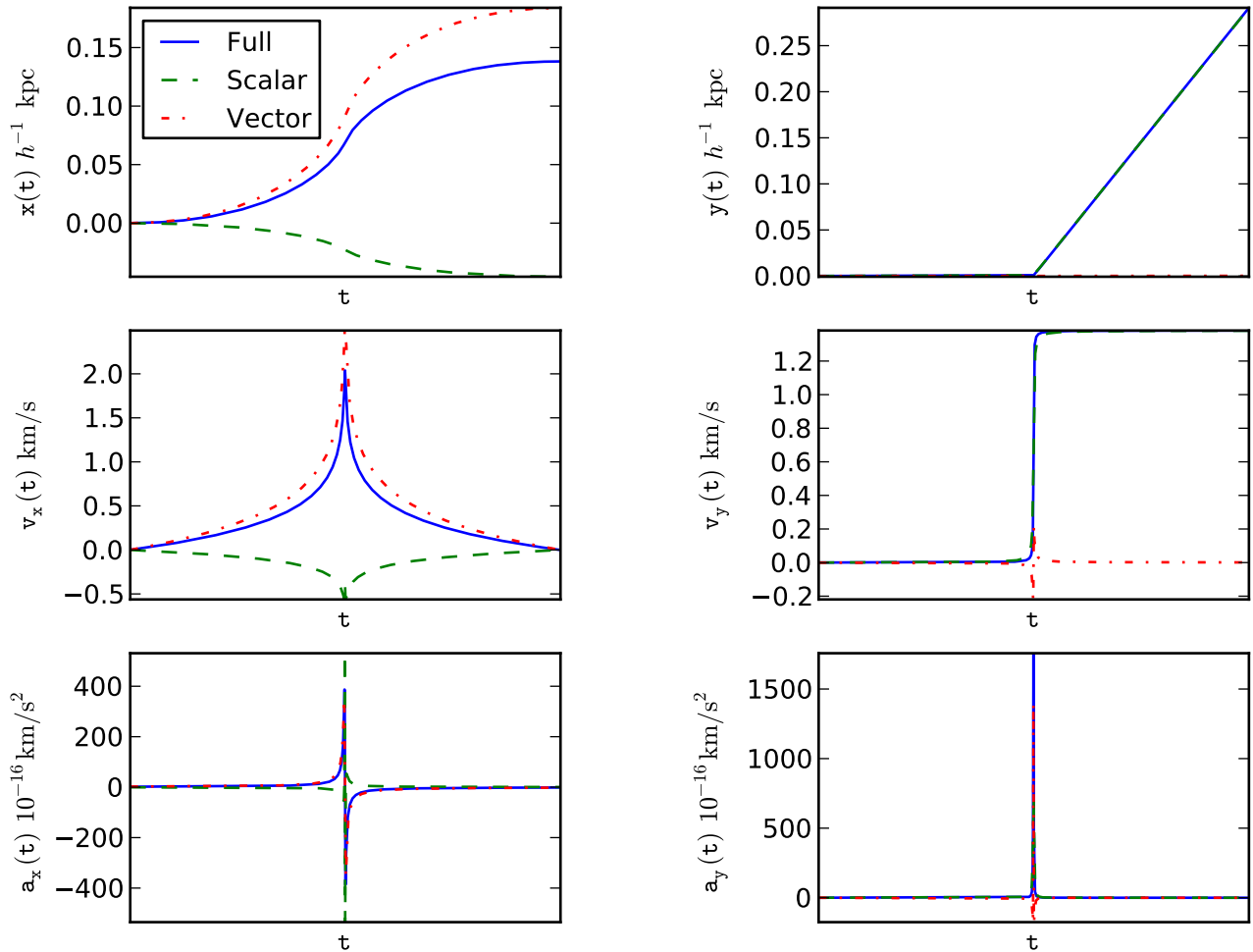}}
\end{center}    
\caption{Motion of a test particle in Minkowski space-time as the string
passes by at a constant velocity $v$, 
from the adapted N-body code results. Initial conditions 
resemble the near by limit case, ie. the initial string particle 
separation is large in the x direction and small in the y direction.
From the top to the bottom the panels show the particle
position, velocity and acceleration. The left-hand panels
show the x-component (along the direction of motion of
the string) and the right-hand panels the y-component
(perpendicular to the direction of motion of the string).}
\label{fig:n-body-results}
\end{figure*}

Details of the particle motion (from the N-body code) are shown
in Fig.~\ref{fig:n-body-results}. The acceleration in the $x$ direction 
(along the motion of the string, left panels of the figure) averages
to zero, so that there is no net velocity left after the string has passed,
and only a small overall displacement. It is however remarkable
how the scalar and vector parts combine to a smooth overall
motion, which is best visible in the middle panel on the left for the
velocity in the $x$ direction. In the $y$ direction (perpendicular to the
string motion) we can see the particle receiving the above-mentioned 
velocity kick. The contribution from the vector part is small and mostly
serves to render the kick more step-like.

\begin{figure}
\begin{center}    
\subfigure{\includegraphics[scale=1]{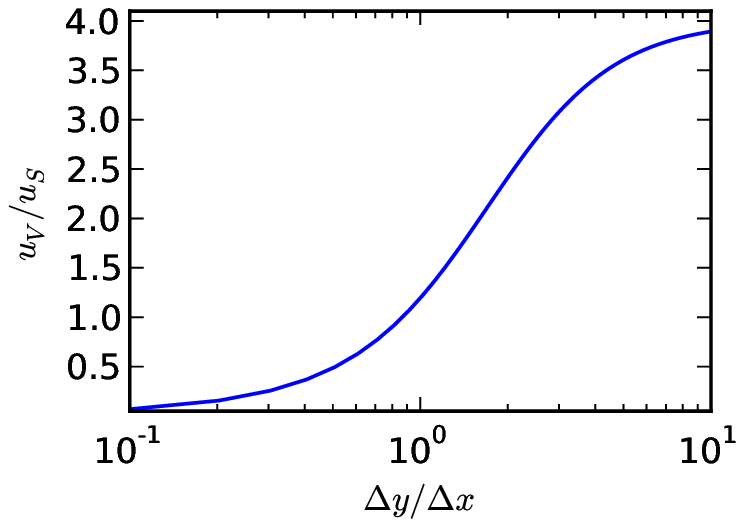}}
\end{center}    
\caption{Ratio of vector to scalar contributions to the final 
velocity in Minkowski space-time. The vector contribution is negligible  in the near by 
limit but dominant if ${\Delta y}/{\Delta x} > 1$.
}
\label{fig:velocity-ratios}
\end{figure}

When looking in more detail at the late-time impact of the scalar and 
vector parts of the acceleration field, we find that in 
the near-by limit the vector part does not contribute 
significantly to the final velocity, see Fig.~\ref{fig:velocity-ratios}. However, 
this is due not least to the special case that we consider, 
where a long, straight string moves on a straight trajectory 
past a particle. In reality we will be dealing with a string 
network, in which case strings are not straight, and they 
move on curved trajectories. In this case we would not expect 
to satisfy the near-by limit at all times. In this situation 
the vector part can contribute at a level comparable to the 
scalar part.

\section{N-body results with a moving straight string in expanding space-time}\label{sec:FLRW}

We now turn to the cosmologically more relevant case of uniformly 
expanding space-time. Specifically, we consider an infinite, 
straight string moving through an initially homogeneous distribution 
of particles expanding with the Hubble flow and follow the subsequent 
evolution of those particles. We present here our leading order results and
put our higher order calculations into Appendix \ref{ap:EOMfirstorderAP}.

We find that the energy momentum tensor $T^{\mu\nu}$ of the 
Nambu Goto string in a FLRW space-time is the same as in Minkowski
space-time, except for a factor $a^{-4}$. This factor cancels out when 
calculating $T_{\mu\nu}$ which is used in the Einstein 
equations (\ref{eq:EINSTEIN},\ref{eq:Tprojectionoperators}) so that
\begin{equation}
    T_{\mu\nu}^{\textrm{FLRW}} = T_{\mu\nu}
\end{equation}


The vector and the tensor perturbations stay the same, while the scalar potentials become

\begin{equation}
\phi(\bv{k},t) = \frac{8\pi^2 G \mu \gamma}{k^2} \left(1- 3 v \dot a \frac{i k_x}{k^2} \right)  e^{i k_x x_s(t)} e^{i k_y y_0} \delta(k_z) 
\label{eq:APsikt2}
\end{equation}
\begin{equation}
\psi(\bv{k},t) = \frac{8\pi^2 G \mu \gamma}{k^2}
\left(v^2 (3\hat{k}_x^2 - 2)+ 3v \dot a \frac{i k_x}{k^2}\right) e^{i k_x x_s(t)} e^{i k_y y_0} \delta(k_z) 
\label{eq:gaugeinvariantvectorfieldFLRW}
\end{equation}

where $x_s(t) = x_0 + 3 v t_0^{2/3} t^{1/3}$. 

$\Sigma_i(\bv{x},t)$ does not change (\ref{eq:sigma_1_x}-\ref{eq:sigma_3_x}) but $\psi(\bv{x},t)$ becomes

\begin{equation}
\psi({\bv{x},t}) =  \frac{G\mu \gamma v}{2r^2} 
\left[  -6 v x(t)^2 + (v -3 \dot a x(t)) r^2
\log \left( \frac{r^2}
{r_0^2} \right)   \right] \label{eq:APsiFLRWXt}
\end{equation}

While the logarithmically divergent contributions at small and large distances in the metric perturbations (\ref{eq:APsiXt}) and (\ref{eq:gaugeinvariantvectorfieldXt}) did not enter the equations of motion in Minkowski space-time, we now have to deal with the term $(\dot{a}/a) \Sigma_1$. 
We discuss in Appendix \ref{app:unphysicallogterm} the (unphysical) origin of these divergences and how we regularise them, and we show that they do not influence the results.

The EOM become

\begin{equation}
\ddot x =  \frac{G \mu \gamma v }{2 a^2 r^4}
\left(
2 v r_x(3r_x^2 + r_y^2) + (14 r_x^2 r^2 - r^4\log{(r^2/r_0^2)})\dot a
\right)
\end{equation}
\begin{equation}
\ddot y = - \frac{G \mu \gamma v r_y}{a^2 r^4}
\left(
v (3 r_x^2 + 5 r_y^2) -  7r_x r^2 \dot a 
\right)
\end{equation}


When the string passes near a particle, 
this particle will be imparted a velocity towards the string (the 
velocity kick discussed above). In physical coordinates with the 
origin fixed to a point on the string trajectory, however, the 
particle is still following the Hubble flow until the recession 
velocity drops below the velocity due to the string passage. The 
particle will then start to move towards the region through which 
the string has passed, and we expect that the particles will
form a wake behind the string once they reach this region. 
Quantitatively, we can compare the particle motion w.r.t. the axis of 
symmetry and the {\em turn-around radius} $r_t$ 
at which the particle motion decouples from the Hubble flow to 
the calculation based on the Zel'dovich approximation in \cite{VilenkinShellardBOOK}:
The physical particle position $y(t)$ is given by
\begin{equation}
y = a \left(y_0 + \xi/a_i\right)
\end{equation}
where $y_0$ is the initial particle position and $\xi(t)$ describes its
displacement,
\begin{equation}
\xi = -\frac{3}{5} u_i t_i \left[\frac{a}{a_i} - \left(\frac{a_i}{a}\right)^{3/2}\right]\epsilon({y_0})
\end{equation}
(at late times, $a\gg a_i$, only the first term in the square brackets is relevant). We set
$\epsilon({y_0}) = 1$ for $y_0 > 0$ and $\epsilon({y_0}) = -1$ for $y_0 < 0$ as
the velocity kick is always towards the string, $u_i$ is the predicted velocity kick 
(\ref{eq:predictedvelocitykick}) and $t_i$, $a_i$ are the 
time and scale factor at which the predicted velocity kick 
occurs (i.e. the moment when $x_{\mathrm{string}} = x_{\mathrm{particle}}$ for a 
particular particle).  From the particle trajectory $y(t)$
it is easy to compute the turn-around radius, since at turn-around $\dot{y}=0$. One finds
that for the particle that turns around at time $t$, 
\begin{equation}
r_t = -\frac{a}{a_i}\xi .
\end{equation}

For testing purposes our simulation starts with a homogeneous distribution 
of $32^3$ particles on an uniform grid in a box size of 
$L=300\,\mathrm{kpc}/h$. We start at redshift $z=99$. The string is initially 
at position $(-10L,L/2)$. We evolve the simulation with the 
string until it reaches position $(10L,L/2)$ at which point 
we turn it off. Thereafter we evolve only the particles and
since we use these simulations to test the implementation of
our formalism, we use the expansion rate of a matter dominated universe
throughout as the approximation above was derived under this assumption.

\begin{figure}
\begin{center}    
\subfigure{\includegraphics[scale=1]{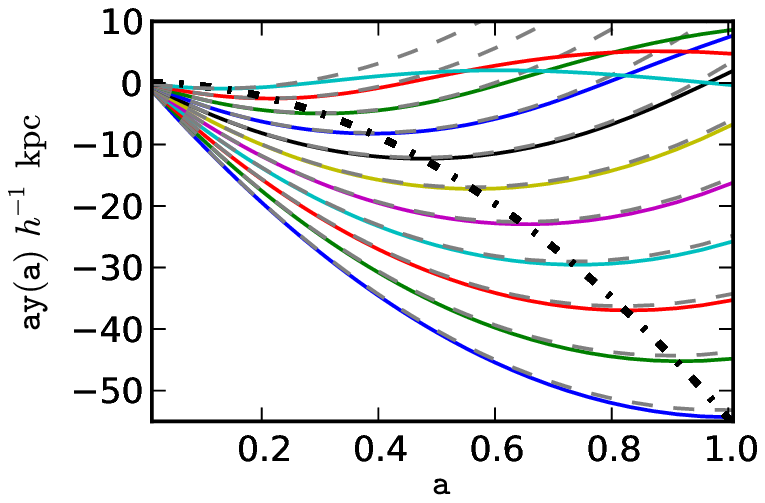}}
\end{center}    
\caption{Particle trajectories vs. scale factor. The 
solid lines are the simulated particle trajectories 
for several sample particles at different initial
distances from the passing string. The dashed lines 
show the corresponding trajectories based on the 
analytical predictions. The thick dash dotted line 
is the analytically-predicted turnaround radius. The 
analytic trajectories approach the axis of symmetry 
faster due to the fact that analytically all particles 
are treated as obeying the near by approximation. Their 
velocity kick is hence larger than the one calculated in 
the simulation. Furthermore, in the simulation we can see
the second turnaround as the particles fall back into the wake.}
\label{fig:r_t-trajectories}
\end{figure}

Our results and comparison to the analytically-predicted
trajectories and turnaround radii $r_t$ are shown in 
Figure~\ref{fig:r_t-trajectories}. The numerically calculated
particle trajectories initially follow closely the analytical 
predictions. Eventually the numerical results diverge from the analytical 
ones due to the fact that in the simulation the velocity
kick for particles that are further away from the axis of symmetry
is smaller than the analytically predicted velocity kick, which is 
derived assuming that the string that came from and went to infinity.

\section{Conclusion and Outlook}

In this paper we set up the formalism needed to include 
topological defects and other sources of weak gravitational 
fields in N-body simulations. For this purpose, we have derived the 
equations of motion of massive particles (Equation \ref{eq:EOMfullyperturbed})
in a perturbed background.  We find that both scalar and vector (gravito-magnetic)
perturbations contribute significantly to leading order, $(\dot x^i/c)^0$. 

Tensor perturbations on the other hand do not contribute to leading order,
and their contribution to the acceleration is suppressed by one power of $\dot x^i/c$
(we investigate the impact of the tensor perturbations in Appendix \ref{ap:EOMfirstorderAP}).

To test and illustrate the equations, we have applied the leading order result 
(Equation \ref{eq:EOMfullyperturbedzerothorder}) to the well known example of a
straight string. We have recovered the expected velocity kick in the limit
where the string passes very close to a particle (relative to the distance that
the string travels) and we have found that the
vector contribution to the final particle motion is sub-dominant. 
Furthermore we have recovered the turn-around radii in FLRW spacetime.

Note that this is the first full numerical calculation (to first order in $\dot x^i$) 
where the velocity kick was not simply taken as an initial condition.

Our results 
can be used for calculating the effect of any sources of weak gravitational
fields on non relativistic and on relativistic particles. 
The focus of our continuing work is on doing large scale N-body 
simulations and calculating the effect of Abelian Higgs cosmic 
string networks \cite{Bevis2007PhRvD..75f5015B} on large scale structures.

\begin{acknowledgments}
We thank Volker Springel for his N-body code Gadget2 \cite{Springel:2005mi}.
This work was partially funded by a STFC DPhil studentship. M.O. and M.K. 
acknowledge financial support by the Swiss National Science Foundation.
I.T.I. was supported by the Southeast Physics Network (SEPnet). 
We acknowledge support from the Science and Technology Facilities Council [grant numbers ST/F002858/1, ST/I000976/1].
\end{acknowledgments}

\appendix

\section{Unphysical $\log({r^2/r_0^2})$ term}\label{app:unphysicallogterm}
An additional complication in the case of expanding background 
compared to the Minkowski space-time is evidenced by 
equation~(\ref{eq:EOMfullyperturbedzerothorder}). Since now
$\dot{a}\neq 0$, there is an additional force term proportional
to $\Sigma_i$, which is logarithmically-diverging with distance
to the string. This unphysical divergence arises because of the 
idealised nature of the infinitely long, straight Nambu-Goto 
string: A string of finite thickness would regularise 
automatically the divergence for small radii, $x^2+y^2\rightarrow 0$,
 while the divergence at large distances, $x^2+y^2\rightarrow\infty$ 
is due to the assumed infinite length of the string. This is
analogous to the logarithmic divergence exhibited by the 
electrostatic potential of an infinite line charge. The 
divergence at small separations is not an issue here because the 
physical thickness of a realistic string is very small compared 
to the typical inter-particle separation in a cosmological 
N-body simulation. However, we would expect that in a 
cosmological setting the size of the causal horizon would 
provide an upper cutoff, above which strings are not formed. 

For this reason we have decided to set $r_0$ to horizon size 
and $r$ such that $\log{(r^2/r_0^2)} \approx 0$ for all particles initially 
(we place the string initially at a distance $\sim r_0$ from the box and 
since $r_0 \gg L$, initially $r \approx r_0$ 
for all particles).


This term only affects the motion in the $x$ direction, 
and hence it does not influence our results and comparison 
to the analytical predictions discussed above.

\section{Tensor Perturbations}\label{ap:tensorperturbations}

To get the contribution due to the tensor term 
we solve the differential equation (\ref{eq:EINSTEIN}) for the tensor source

\begin{equation}
\tau_{ij}^{(\pi)} = \pi\gamma\mu(1-v^2 \hat k_x^2)
N_{ij}
e^{i k_x (x_0 + v\tau)} e^{i k_y y_0} \delta(k_z),
\label{eq:gaugeinvarianttensorfield}
\end{equation}

where 
\begin{equation}
N_{ij} =
\left(
\begin{array}{cccc}
\hat k_y^2 & -\hat k_x \hat k_y & 0 \\
-\hat k_x \hat k_y & \hat k_x^2 & 0 \\
0 & 0 & -1
\end{array}
\right).\nonumber
\end{equation}

For a general tensor sorce, the numerical method involves solving the differential
equation on each grid point in Fourier space, and then numerically inverse Fourier 
transforming the resulting $\dot h_{ij}$ which is used in the equation of motion.

However, in the case of the Nambu Goto string we directly solve the differential equation analytically
for modes that are well inside the horizon, ie $k\tau \gg 1$. Dropping the friction term 
(note that this term also disappears in Minkowskian spacetime) we can rewrite it
to read

\begin{equation}
\dprime h^{(T)}_{ij} + k^2 h^{(T)}_{ij} = S e^{i k_x v \tau}
\end{equation}

where we define and use $S e^{i k_x v \tau} = 8\pi G\tau_{ij}^{(\pi)}$ to make 
the time dependence explicit. Hence

\begin{equation}\label{eq:solwithh0}
  h^{(T)}_{ij} = \frac{S e^{i k_x v \tau }}{k^2 - k_x^2v^2}  + h^{(T)}_{ij,0}
\end{equation}

where $h^{(T)}_{ij,0}$ is a solution to the homogeneous equation 
$\dprime h^{(T)}_{ij} + k^2 h^{(T)}_{ij} = 0$, determined by the initial conditions.

Equation (\ref{eq:solwithh0}) represents the tensor part of the boosted static 
gravitational field of the string when $h^{(T)}_{ij,0} = 0$. 
Hence we arrive at our solution

\begin{eqnarray}
  h^{(T)}_{ij} &=& \frac{S e^{i k_x v \tau }}{k^2 - k_x^2v^2}  \\
  \acute h^{(T)}_{ij} &=& i k_x v\frac{S e^{i k_x v \tau }}{k^2 - k_x^2v^2} \label{eq:hdot_k}
\end{eqnarray}

We also find the analytic solution to the full differential 
equation (\ref{eq:EINSTEIN}) for the tensor source (\ref{eq:gaugeinvarianttensorfield})

\begin{eqnarray}
\tilde h^{(T)}_{ij} &=& \frac{S e^{i k_x v \tau }}{(k^2 - k_x^2v^2)^3 \tau^3}
\Big(-8 i k_x v - 8 k_x^2 v^2 \tau
\\&&  \nonumber
 + 4 i k_x v (-k^2 + k_x^2v^2)\tau^2 + (k^2 - k_x^2v^2)^2\tau^3\Big) ~.
\end{eqnarray}

However, we find that $1-|\tilde h^{(T)}_{ij}/h^{(T)}_{ij}| \approx 10^{-15}$
when comparing over our ranges and scales of interest ($0.01 < a < 1$, $2\pi/L < k_i< 512\pi/L$)
for which $1.7\times10^4 < k\tau < 4.4\times10^4$.


\section{Equations of motion to first order in the velocity}\label{ap:EOMfirstorderAP}

The equation of motion to first order in the particle velocities, in physical time, is

\begin{eqnarray}\label{eq:EOMfullyperturbedfirstorder}
\ddot x^i 
+ 2\frac{\dot a}{a} \dot x^i
&=& 
-\frac{1}{a^2}\partial_i\psi 
+ \frac{1}{a}\dot \Sigma_i 
+ \frac{\dot a}{a^2} \Sigma_i
+ ( \dot\psi - 2\dot\phi) \dot x^i\nonumber
\\&&
+ \frac{1}{a}(\partial_j\Sigma_i - \partial_i\Sigma_j)\dot x^j
- \dot h_{ij}^{(T)}\dot x^j
\end{eqnarray}

\begin{figure*}
\begin{center}    
\subfigure{\includegraphics[scale=1]{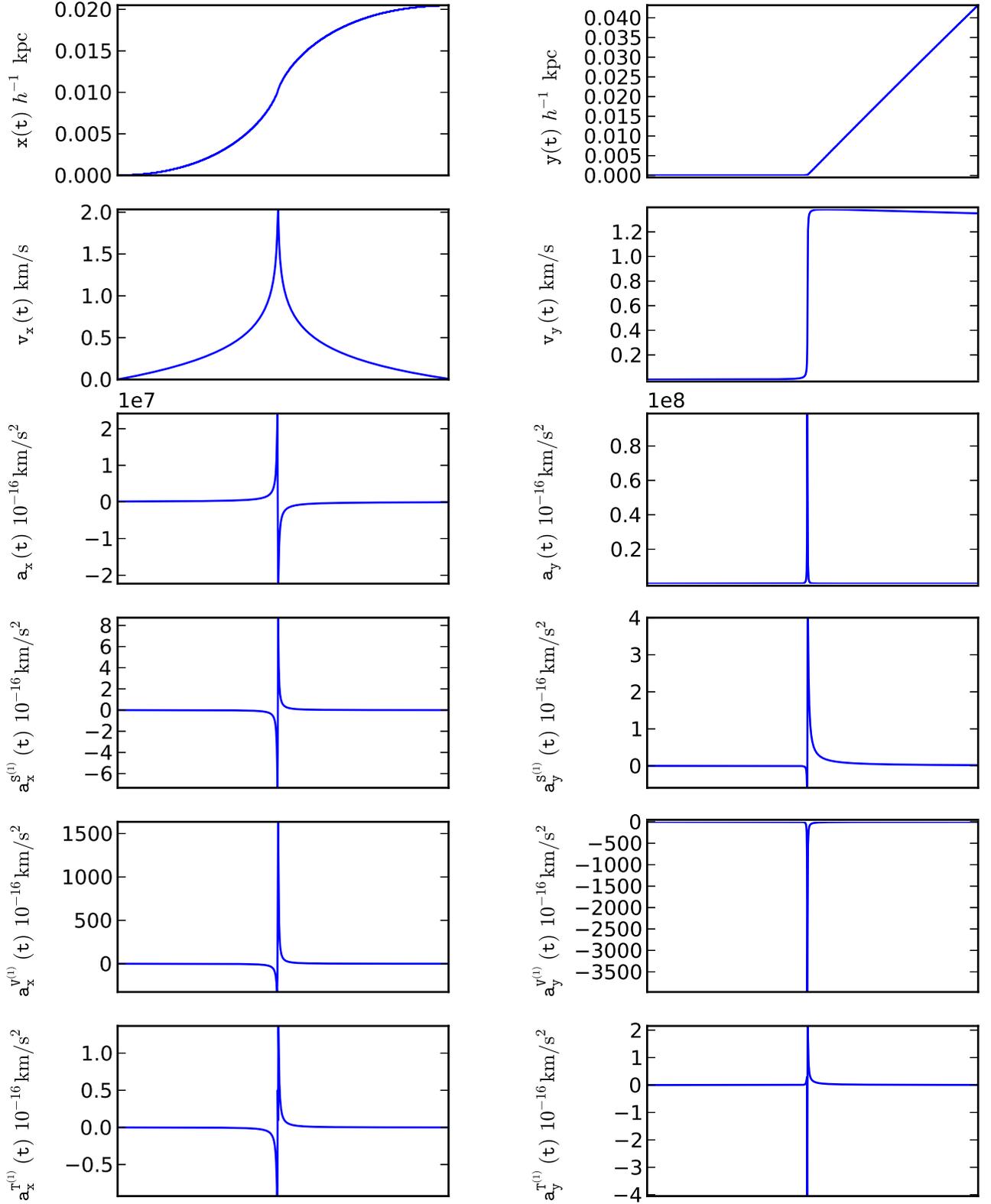}}
\end{center}    
\caption{Motion of a test particle in FLRW space-time with the same initial 
conditions as in section \ref{sec:FLRW}. 
From the top to the bottom the panels show the comoving particle
position, physical velocity and comoving accelerations. 
The third row shows the total comoving acceleration and the rows below
show the first order scalar, vector and tensor contributions to it.}
\label{fig:n-body-1st-results}
\end{figure*}

We find that all the contributions to first order in the particle 
velocity are significantly smaller (suppressed by one power of $(\dot x^i/c)$) 
and they hence do not influence the result (see Figure~\ref{fig:n-body-1st-results}).
For completeness we list here all the additional first order analytical results. 

The first order scalar contributions are given by

\begin{eqnarray}
\dot \psi - 2 \dot \phi &=& \frac{G \mu \gamma}{2 r^4} \nonumber
\Big\{
(v - \dot x)
\Big(
-2 r_x \Big[4 r^2 + v^2(-5 r^2 + 6 r_x^2)\Big] 
\\&&
+ 9 r^2 v \dot a \Big[2 r_x^2 + r^2 \log{(r^2/r_0^2)} \Big]
\Big)  \\ \nonumber
&&
+\; 2 r_y \dot y \Big[4 r^2 + v^2(r^2 + 6 r_x^2) - 9 r^2 v r_x \dot a\Big] 
\Big\}
\end{eqnarray}

The first order vector contributions are given by

\begin{equation}
\partial_j \Sigma_1 \dot x^j = - 4 G \mu \gamma v \left[r_x \dot x  (r_x^2 - r_y^2) + r_y \dot y  (3r_x^2 + r_y^2)\right]/r^4
\end{equation}
\begin{equation}
\partial_j \Sigma_2 \dot x^j = - 4 G \mu \gamma v \left[(r_x^2 - r_y^2)(r_x \dot x  + r_y \dot y )\right]/r^4
\end{equation}
\begin{equation}
(\partial_j\Sigma_1 - \partial_1\Sigma_j)\dot x^j = -8 G \mu \gamma v r_y \dot y / r^2 
\end{equation}
\begin{equation}
(\partial_j\Sigma_2 - \partial_2\Sigma_j)\dot x^j = 8 G \mu \gamma v r_y \dot x / r^2
\end{equation}

Note that the first two of these terms come from the 
$\dot \Sigma_i \equiv (d/dt)\Sigma_i(t,x(t),y(t))$ term.

The first order tensor contributions are given by inverse Fourier transforming Equation (\ref{eq:hdot_k})

\begin{eqnarray}
\dot h_{11}^{(T)} &=& - G  \mu  \gamma v r_x(-r_x^2+ r_y^2)/r^4 \\
\dot h_{22}^{(T)} &=&   G  \mu  \gamma v r_x( r_x^2+3r_y^2)/r^4 \\
\dot h_{33}^{(T)} &=& -2 G \mu  \gamma v r_x/r^2 \\
\dot h_{21}^{(T)} &=&  G   \mu  \gamma v r_x(r_x^2 - r_y^2)/r^4
\end{eqnarray}

Finally, note that in equation (\ref{eq:EOMfullyperturbed}) there are further terms
of order $(\dot x^i/c)^2$ and $(\dot x^i/c)^3$ which will be smaller yet.

\bibliography{citations}

\end{document}